# Generation of Spectral Clusters in a Mixture of Noble and Raman-Active Gases

Pooria Hosseini,[1,*] Amir Abdolvand,[1] and Philip St.J. Russell[1,2]

[1]*Max Planck Institute for the Science of Light, Geunther-Scharowsky-Str. 1, 91058 Erlangen, Germany*
[2]*Department of Physics, University of Erlangen-Nuremberg, 91058 Erlangen, Germany*
*\*Corresponding author:* pooria.hosseini@mpl.mpg.de

We report a novel scheme for the generation of dense clusters of Raman sidebands. The scheme uses a broadband-guiding hollow-core photonic crystal fiber (HC-PCF) filled with a mixture of $H_2$, $D_2$, and Xe for efficient interaction between the gas mixture and a green laser pump pulse (532 nm, 1 ns) of only 5 µJ energy. This results in the generation from noise of more than 135 ro-vibrational Raman sidebands covering the visible spectral region with an average spacing of only 2 THz. Such a spectrally dense and compact fiber-based source is ideal for applications where closely spaced narrow-band laser lines with high spectral power density are required, such as in spectroscopy and sensing. When the HC-PCF is filled with a $H_2$-$D_2$ mixture the Raman comb spans the spectral region from the deep UV (280 nm) to the near infrared (1000 nm).

The emergence over the last 20 years of hollow-core photonic crystal fiber (HC-PCF) [1] has caused a revival of interest in the field of gas-based nonlinear optics [2]. For example, the generation and study of the coherence properties of noise-seeded, multi-octave Raman combs generated by stimulated Raman scattering (SRS) in gas-filled HC-PCFs has led to several breakthroughs [3-5]. Due to the long interaction lengths and tight confinement of light in these fibers, it is possible, using nanosecond pulses with energies of only a few µJ, to achieve high values of net Raman gain $G_{ss} = g_{ss} I_p L$ (where $g_{ss}$ is the steady-state Raman gain coefficient, $I_p$ the pump intensity, and $L$ the fiber length) while operating in the transient regime, that is when the pump pulse duration $t_p < G_{ss}/\Gamma$, where $\Gamma$ is the damping rate of the Raman coherence [6]. These conditions allow, for each pump pulse, the generation from noise of a Raman comb with spectral components that have strong mutual coherence [4, 5]. With proper tuning of the spectral phase, e.g., by using a spatial light modulator (SLM), this can result in the generation of a (sub-) femtosecond pulse train within each single pump pulse [7]. Motivated by these possibilities, many studies to date have concentrated on increasing the spectral width of the Raman comb using novel HC-PCFs [3, 4, 8]. These studies make use of synchronous molecular oscillations at the Raman frequency $v_R$ to phase-modulate an incident pump pulse, creating Raman sidebands spaced $v_R$ apart in the frequency domain. To synthesize ultrashort pulses, a high Raman frequency (e.g., the 125 THz vibrational transition in hydrogen) is desirable. This has, however, the drawback that the repetition rate of transform-limited pulses will equal the Raman frequency, which is unmanageably high. An additional challenge is utilizing the full width of the spectrum, which can extend from the vacuum UV to the infrared and require ultrabroadband achromatic performance from both the optics and the SLM. It may therefore be desirable to have a source with a finer comb spacing over a less broad frequency interval.

Sources of spectrally dense but discrete narrowband laser lines are also of great importance in spectroscopy [9]. Yavuz et al. have shown how, using $H_2$ and $D_2$ as two independent molecular phase modulators in series offering a total of two Raman modes or modulation frequencies, it is possible to increase the spectral density of the Raman sidebands and thus decrease the repetition rate of the resulting pulse train [10].

Since their scheme relied, however, on using three lasers with frequencies tuned to two different Raman transitions, only two Raman modes could be efficiently excited.

Early studies on SRS in a mixture of Raman active and noble gases in free space or conventional gas cells have shown that the addition of a dispersive noble buffer gas (NBG) increases the SRS phase-mismatch to higher-order Stokes/anti-Stokes sidebands, resulting in preferential conversion of the pump to the first few Stokes lines [11-14]. These studies have also reported that the conversion efficiency is impaired by a reduction in Raman gain as the NBG pressure is increased. Despite these limitations, mixtures of $H_2$, $D_2$, and $CH_4$ in various NBGs were explored, the aim being to optimize the conversion efficiency to few selected Stokes sidebands. These experiments were mostly performed in the continuous wave regime, when the Raman gain falls as the pressure of the buffer gas is increased.

Here we show that by working in the transient stimulated Raman scattering (TSRS) regime the gain is only marginally reduced by addition of a NBG, allowing highly efficient Raman frequency convertors to be constructed. We use a gas-filled HC-PCF, which offers long interaction lengths and tight field confinement, making it possible to reach high net Raman gain even in the transient regime.

A schematic of the experimental set-up is shown in Fig. 1(d). A Q-switched laser delivering 1 ns pulses at 532 nm at a repetition of 800 Hz is used as the pump source. A quarter wave-plate converts the polarization of the pump from linear to circular, thus enhancing the rotational Raman gain and the strength of the rotational Raman sidebands [15]. The pump is coupled into a three meters of a kagomé HC-PCF with a core diameter of 24 μm and 1 dB/m loss at 532 nm. The fiber is filled with a variety of mixtures of $D_2$, $H_2$, and Xe. The spectrum is imaged on a screen by dispersing it using a prism and is recorded using a spectrometer.

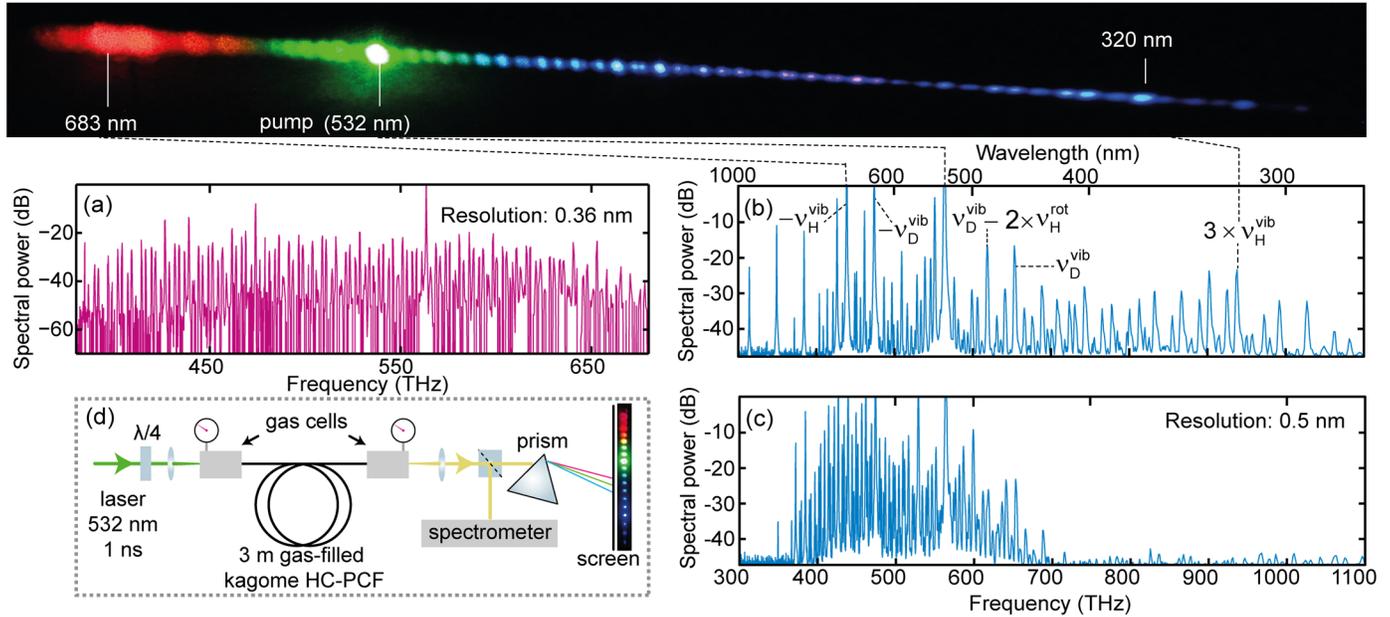

Fig. 1 (a) A flat spectral cluster covering the visible-UV range and measured by a high-resolution OSA (0.36 nm). More than 135 lines are resolvable at −40 dB. The cluster is generated in a $H_2$-$D_2$-Xe mixture, with respective partial pressures of 5-8-4 bar, when pumping the fiber with 5 μJ pulses at 532 nm. (b) The Raman comb generated in a $H_2$-$D_2$ mixture in the absence of Xe and measured by an Ocean Optics spectrometer (resolution 0.5 nm). All the other experimental conditions are as in (a). The positions of some of the combination lines are indicated (see the text for a definition of the notation). The top panel shows the spectrum in (b) when dispersed by a prism and captured on a screen. (c) The spectrum after the addition of 4 bar of Xe, as in (a) but captured with the spectrometer. Note how the spectrum is clustered in the visible with increased spectral density. (d) A schematic of the experimental set-up.

By exploiting the unique properties of gas-filled HC-PCF, a spectral cluster of more than 135 Raman sidebands with an average spacing of only 2 THz in the visible is generated in a H$_2$-D$_2$-Xe mixture using single 532 nm pump pulses with 5 µJ energy (Figs. 1(a) and 1(c)). This is the densest Raman comb ever generated using molecular modulation. When the HC-PCF is filled with a H$_2$-D$_2$ mixture, a Raman comb spanning from the deep UV (280 nm) to infrared (1000 nm) is also generated, as may be seen in Fig. 1(b) (also see the photograph of the dispersed spectrum in the top panel).

Addition of Xe as a NBG has the advantage of allowing the dispersion of the "gas + fiber" system to be tailored, while leaving the net Raman gain almost unaffected. This is because, in the highly transient regime, the gain does not undergo a notable decrease due the increasing Raman gain linewidth ($\Delta v$ [s$^{-1}$] = $\Gamma/\pi$), caused by the collisions with the molecules of the buffer gas. For a two-component gas mixture the dependence of the Raman gain linewidth on the partial pressures of the gas constituents is given by [12]:

$$\Delta v = A/\rho_G + B\rho_G + C\rho_B, \quad (1)$$

where $\rho_G$ and $\rho_B$ are respectively the partial number densities of the main Raman and buffer gases in amagat, where $\rho$ [amagat] corresponds to the ratio of the molecular number density $n$ ($p$, $T$) to the number density at standard pressure and temperature $n_0$ ($p_0$, $T_0$). The values of the broadening coefficients $A$, $B$, and $C$ for vibrational and rotational transitions in the presence of Xe and Ar buffer gases are shown in Table 1. Note that for some transitions data for $C$ coefficients is not available in the literature.

According to the theory of SRS [6] in the continuous wave or steady state (SS) regime, the Stokes intensity grows exponentially with a gain factor $G_{SS}$ that is inversely proportional to the Raman linewidth, i.e., $G_{SS} \propto \Gamma^{-1}$. In the absence of Dicke narrowing (above 10 bar of H$_2$ for vibrational Stokes), the addition of a NBG introduces extra line-broadening through additional collisions so that the SS Raman gain will drop as the pressure of the buffer gas increases. In the transient regime, however, the Stokes intensity grows to a very good approximation with a gain factor [6]:

$$G_T \approx \sqrt{8G_{SS}t_p\Gamma} - 2t_p\Gamma < G_{SS}. \quad (2)$$

It may be seen that for large values of net Raman gain $G_{SS}$, the second term on the right-hand side of Eq. 2 can be neglected. In this regime the product $G_{SS}\Gamma$ is therefore independent of the linewidth and $G_T$ is only weakly influenced by the introduction of a NBG. This is true provided the system is still operating in the transient regime. Increasing the linewidth, and thus the dephasing rate, further will push the system towards SS-like behavior, i.e., the gain will decrease with further increase in NBG pressure.

**Table 1. Raman linewidth self-broadening coefficients $A$ and $B$ for H$_2$ and D$_2$, and in the presence of a noble buffer gas, $C$ [12, 16].**

| Transition | A | B | $C_{Xe}$ | $C_{Ar}$ |
|---|---|---|---|---|
| | MHz× amagat | MHz/amagat | | |
| vib. H$_2$ | 309 | 51.8 | 414 | 256.2 |
| vib. D$_2$ | 101.2 | 116.7 | … | 318 |
| rot. H$_2$ | 6.15 | 114 | … | 93.9 |
| rot. D$_2$ | 1.94 | 160 | … | 74.3 |

To verify that the system operates in the transient regime even in the presence of Xe, we filled the fiber with 3 bar of Xe and 10 bar of H$_2$ and recorded the temporal profiles of the pump and the first vibrational Stokes for a pump energy of 0.3 µJ using a fast photodiode and an oscilloscope. The input energy was kept intentionally low so as to exclude the generation of the higher order Stokes/anti-Stokes light. Under these conditions the Raman gain coefficient takes the value of g$_{ss}$ = 0.77 cm/GW, which is 30% of its value without the buffer gas (2.46 cm/GW) [16]. Nevertheless, the condition for TSRS is still fulfilled ($t_p < G_{SS}/\Gamma$ = 12 ns). The recorded temporal profiles are plotted in Fig. 2. It can be clearly seen that the Stokes pulse is shorter in duration than the initial pump pulse, its peak being delayed by almost half a nanosecond with respect to the peak of the pump. This clear evidence that the system is operating in the TSRS regime [17].

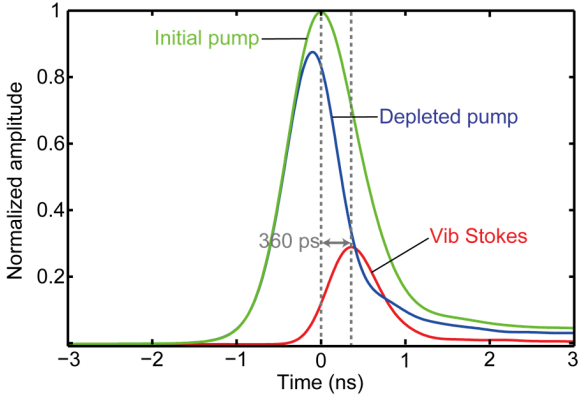

Fig. 2 Temporal profiles of the initial pump pulse (green, 300 nJ), the output pump (blue) and the first vibrational Stokes (red) in a mixture of 3 bar Xe and 10 bar H$_2$. The delay of ~400 ps between the peak of the Stokes and the initial pump pulse indicates operation in the transient regime.

To investigate further the effect on the Raman response of adding a NBG, we performed second experiment in which a Raman comb was generate in 10 bar of H$_2$ while gradually adding Xe. With pulse energies of 5 µJ, a high net Raman gain of $G_{SS} \sim 1700$ is achieved, which relaxes the condition for TSRS, even with addition of 12 bar of Xe. A ro-vibrational comb extending from 270 nm to 1100 nm with more than 45 sidebands was generated in pure H$_2$ ($\rho_{Xe} = 0$). Figure 3 shows a color-coded graph of the spectra recorded when the partial pressure of Xe was varied from 0 to 12 bar. All the other experimental parameters, e.g., the pump power and H$_2$ partial pressure, were kept constant throughout the experiment. As the pressure of the buffer gas increased, the anti-Stokes and higher order Stokes lines became weaker. At 10 bar of Xe, most of the energy is concentrated in the spectral vicinity of the pump, predominantly on the Stokes side, while the residual pump energy shows only a 2 dB increase at maximum. This strongly indicates that the transient Raman gain does not decrease significantly when Xe is added. The observed reduction in the spectral spread of the comb-lines at higher Xe pressure is mainly due to the increased phase mismatch for higher order anti-Stokes lines, which is more pronounced at higher frequencies. This unique behavior, which leaves the Raman gain basically unchanged while the phase-mismatch can be controlled by tuning the pressure of the NBG, makes it possible to generate a comb of ro-vibrational sidebands with high spectral power over a narrow frequency range. This is illustrated in the inset of Fig. 3 for 12 bar of Xe. The main part of the spectrum, which contains most of the energy, consists of the pump, the first six rotational Stokes sidebands and the first vibrational Stokes. The spectrum is flat to within ~3 dB and all the sidebands have higher spectral power than when no buffer gas is present.

To understand how this works, we define the frequency shift of the (m, n)-th ro-vibrational line as $v_{comb} - v_0 = m v_H^{vib} + n v_H^{rot}$, where $v_0$ is the pump frequency, $v_H^{vib(rot)}$ the vibrational (rotational) Raman frequency shift and m and n are integers. Since in our case the dispersion is such that parametric gain suppression does not occur [18], cascaded, phase-matched generation of higher-order Raman sidebands will rely on minimizing the linear phase mismatch. For purely vibrational transitions (n = 0) the dephasing rate is given by $\Delta \beta_{m,0} = \beta_{m,0} - \beta_{m-1,0} + (\beta_{0,0} - \beta_{-1,0})$, where $\beta_{m,0}$ is the propagation constant of the mth vibrational component and $\beta_{0,0} - \beta_{-1,0}$ is the wavevector of the coherence wave created by interference of the pump and the first vibrational Stokes, which drives the SRS process. Figure 4 shows the dispersion curves for a H$_2$-Xe gas mixture with 10 bar partial pressure of H$_2$ and different partial pressures of Xe [2, 19]. The dephasing rate ($\beta_{0,0} - \beta_{-1,0}$) increases for the higher Stokes and anti-Stokes lines as the partial pressure of Xe increases, which agrees with the observation in Fig. 3. This concentrates the pump energy in the spectral region between the pump and the first few vibrational Stokes lines, resulting in strong comb-lines. This is also true for the rotational Stokes lines ($v_H^{rot} = 18$ THz), which occupy the frequency interval between the pump and the first vibrational Stokes.

If Ar is used in place of Xe, a NBG partial pressure of 60 bar is needed to achieve a similar dispersion landscape. This results in a 3 times broader linewidth (see Table 1), which considerably reduces the transient Raman gain due to the linear term in Eq. 1 [12].

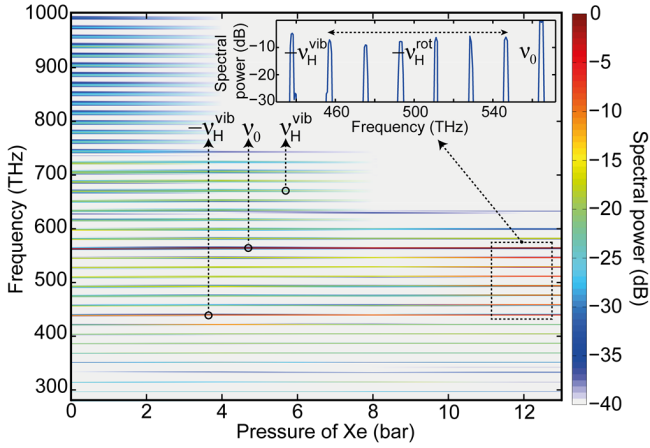

Fig. 3: Evolution of the Raman comb for a fixed partial pressure of $H_2$ (10 bar) as a function of the Xe partial pressure. The pump energy is fixed at 5 µJ. The position of the pump and first vibrational Stokes/anti-Stokes lines are indicated. The short wavelength edge is more sensitive to the increase in Xe pressure, which results in an increased phase-mismatch. Inset shows a cut through the map at ~12 bar of Xe. Spectrum is flat within 3 dB.

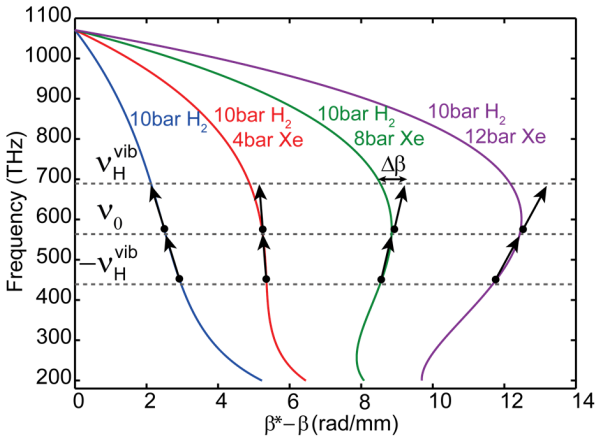

Fig. 4 Dispersion curves for a kagomé-PCF with a core diameter 24 µm, filled with 10 bar partial pressure of $H_2$ (blue curve). The red, green, and violet curves are for Xe partial pressures of 4, 8, and 12 bar. The parameter $\beta^* = (\omega/\omega_{max})\beta(\omega_{max})$ with $\omega_{max} = 1070$ THz, is used to magnify the weak dispersion changes. The black solid arrows indicate the coherence waves that drive SRS. The dashed lines mark the position of the pump, first vibrational Stokes and anti-Stokes. $\Delta\beta_{1,0}$ is the dephasing parameter for the anti-Stokes line, taking the values 0.033, 0.34, 0.72, and 1.10 (rad/mm) for the four Xe pressures shown.

In order to increase the spectral density we used a mixture of $H_2$-$D_2$, with strongest vibrational and rotational Raman shifts of $\nu_H^{vib} = 4155$ cm$^{-1}$ (125 THz), $\nu_H^{rot} = 587$ cm$^{-1}$ (18 THz) and $\nu_D^{vib} = 2987$ cm$^{-1}$ (90 THz), and $\nu_D^{rot} = 415$ cm$^{-1}$ (12.5 THz). The frequency shifts offered by such a mixture are $m\nu_H^{vib} + n\nu_H^{rot} + k\nu_D^{vib} + l\nu_D^{rot}$ where $(m, n, k, l)$ are integers [20] (see Fig. 1).

We are now in a position to discuss the measurement in the top panel of Fig. 1 in more detail. It was made for partial pressures of 5 bar ($H_2$) and 8 bar ($D_2$), a choice that is based on requiring the transient vibrational Raman gain coefficients (Eq. 1) of each gas to be similar. This improves the spectral flatness of the comb as well as increasing the generation efficiency of the combination lines [21]. The choice of total pressure is less critical, although one needs to balance the overall Raman gain against the increase in dephasing to the anti-Stokes sidebands at high pressures. The results show a quasi-continuous comb extending from the IR (1000 nm) to the UV (280 nm). The dashed lines in Fig. 1(b) indicate the positions of the pump as well as several of the frequency-shift combinations $\nu_{comb} - \nu_0 = m \times \nu_H^{vib} + n \times \nu_H^{rot} + k \times \nu_D^{vib} + l \times \nu_D^{rot}$. Figure 1(c) shows the spectrum when 4 bar of xenon is added to the mixture. As discussed above, increasing the dephasing to the higher order Stokes and anti-Stokes lines restricts comb generation to a narrower spectral region, in this case ~350-700 THz, while at the same time considerably increasing the spectral density of the Raman comb. Indeed, the spectrometer is unable to resolve all the lines (note the pedestal in Fig. 1(c)).

Figure 1(a) shows the same spectrum recorded using an optical spectrum analyzer (OSA) with a resolution of 0.36 nm. More than 135 lines were resolvable down to the –40 dB level with an overall flatness of ~5 dB (with the exception of a few strong vibrational Raman lines). We estimate the average spectral power density of the comb to be 0.1 mW/nm at a pump power of only 4 mW and a pump linewidth of ~100 GHz. This can easily be increased by orders of magnitude, for example, by using a laser with a narrower linewidth or higher average power, or by using a longer fiber.

In conclusion, a dense two-octave spanning UV to IR Raman comb can be generated in a mixture of $H_2$/$D_2$ through TSRS in kagomé-PCF. By tailoring the phase-matching condition and restricting the spectral extent of the comb, a large number of comb-lines can be generated in the visible. Such a spectrally flat and dense comb is ideal for spectroscopy, as well as for synthesizing ultrashort pulses. This approach can easily be adapted for use in other spectral regions, e.g., in the ultraviolet [22]. Further possibilities include pumping the system with high power, picosecond pulses as well as investigating the mutual coherence of the interleaved comb-lines. If present, mutual coherence would permit generation of quasi-periodic waveforms [10].